\documentclass[sigconf]{acmart}

\usepackage{amsmath,amsfonts}
\usepackage{algorithm}
\usepackage{algpseudocode}
\usepackage{graphicx}
\usepackage{textcomp}
\usepackage[disable]{todonotes}
\usepackage{regexpatch}
\usepackage{subcaption}
\usepackage{multirow}
\usepackage{xspace}
\usepackage{orcidlink}
\usepackage{doi}
\usepackage{nccmath}
\usepackage{xspace}
\usepackage{xargs}
\usepackage{microtype}
\usepackage{tabularx}
\usepackage{booktabs}
\usepackage{listings}
\usepackage{parcolumns}
\usepackage[many]{tcolorbox}
\usepackage{float}
\usepackage{soul}
\usepackage{footmisc}
\usepackage{xcolor}
\usepackage{colortbl}
\usepackage{siunitx}
\usepackage{pbalance}

\usepackage[absolute]{textpos}

\definecolor{eclipsegreen}{rgb}{0.32, 0.5, 0.39}
\definecolor{eclipsepurple}{rgb}{0.41, 0.13, 0.32}
\definecolor{eclipsebrown}{rgb}{0.39, 0.25, 0.25}
\definecolor{eclipseblue}{rgb}{0.16, 0.07, 0.72}

\definecolor{matchgray}{gray}{0.92} 
\definecolor{add}{HTML}{E6FFED} 
\definecolor{del}{HTML}{FFEBE9} 
\definecolor{mod}{HTML}{FFF8C5} 
\newcommand{\match}{\rowcolor{gray!15}}

\makeatletter
\xpatchcmd{\@todo}{\setkeys{todonotes}{#1}}{\setkeys{todonotes}{inline,#1}}{}{}

\newcommand{\todoRobin}[1]{\todo[backgroundcolor=blue!20]{@Robin: #1}}

\newcommand{\contribution}[1]{\hyperref[c#1]{\textbf{C#1}}}
\newcommand{\mref}[1]{\hyperref[eval-m]{\textbf{M#1}}}
\newcommand{\qref}[1]{\hyperref[eval-q]{\textbf{Q#1}}}
\newcommand{\summaryBox}[2]{\begin{tcolorbox}[colback=black!1!white,colframe=white!20!black,center,valign=top,halign=left,before skip=2pt,after skip=2pt,center title,sharp corners,boxrule=0pt,boxsep=-2pt,left=5pt,right=5pt,width=\linewidth]\textbf{Answer to \qref{#1}:} \textit{#2}\end{tcolorbox}}

\newcommand{\plotsqueeze}{\vspace{-9pt}}
\newcommand{\figuresqueeze}{\vspace{-8pt}}
\newcommand{\tablesqueeze}{\vspace{-10pt}}

\newcommand{\hugCaption}{\vspace{-17pt}}
\newcommand{\hugPlotCaption}{\vspace{-18pt}}
\newcommand{\groupHugPlotCaption}{\vspace{-15pt}}
\newcommand{\hugTableCaption}{\vspace{-7pt}}

\newboolean{showdiff}
\setboolean{showdiff}{false}   

\ifthenelse{\boolean{showdiff}}{
\emergencystretch 3em
\newcommand{\add}[1]{{\color{blue}#1\color{black}}}
\newcommand{\del}[1]{{\color{red}\st{#1}}}
}{
\newcommand{\add}[1]{#1}
\newcommand{\del}[1]{}
}
\newcommand{\repl}[2]{\del{#1}\add{#2}}

\usepackage[inline]{enumitem}
\usepackage{minted}
\usepackage[normalem]{ulem}

\RequirePackage[nolist,nohyperlinks]{acronym}
\begin{acronym}
\acro{CPG}[CPG]{Code Property Graph}
\acroplural{CPG}[CPGs]{Code Property Graphs}
\acro{EOG}[EOG]{Evaluation Order Graph}
\acro{AST}[AST]{Abstract Syntax Tree}
\acro{DSL}[DSL]{domain-specific language}
\acro{DFG}[DFG]{Data-Flow Graph}
\acro{CFG}[CFG]{Control-Flow Graph}
\acro{DDG}[DDG]{Data Dependency Graph}
\acro{CDG}[CDG]{Control Dependency Graph}
\acro{SMM}[SMM]{Subsequence Match Merging}
\acro{TSN}[TSN]{Token Sequence Normalization}
\acro{Nocte}[\textsc{Nocte}]{NOrmalization-driven Code Transformation Engine}
\end{acronym}

\AtBeginDocument{%
  }

\setcopyright{acmlicensed}
\copyrightyear{2026}
\acmYear{2026}
\setcopyright{cc}
\setcctype{by}
\acmConference[ICSE '26]{2026 IEEE/ACM 48th International Conference on Software Engineering}{April 12--18, 2026}{Rio de Janeiro, Brazil}
\acmBooktitle{2026 IEEE/ACM 48th International Conference on Software Engineering (ICSE '26), April 12--18, 2026, Rio de Janeiro, Brazil}
\acmDOI{10.1145/3744916.3773225}
\acmISBN{979-8-4007-2025-3/26/04}

\usepackage{csquotes} 

\usepackage[many]{tcolorbox}

\usepackage[many]{tcolorbox}

\begin{document}
\begin{textblock}{14}(1,0.5)
\noindent \begin{center}\textbf{This is the author's version of the paper submitted and accepted at ICSE'26.\\The definitive Version of Record is published here: \url{https://doi.org/10.1145/3744916.3773225}.}
\end{center}
\end{textblock}

\title{Same Same But Different: Preventing Refactoring Attacks on Software Plagiarism Detection}

\author{Robin Maisch}
 \email{robin.maisch@kit.edu}
 \orcid{0009-0003-4546-9048}
 \affiliation{%
   \institution{KASTEL -- Institute of Information Security and Dependability, Karlsruhe Institute of Technology (KIT)}
   \city{Karlsruhe}
   \country{Germany}
}

\author{Larissa Schmid}
\authornote{Work was conducted while affiliated with KASTEL - Institute of Information Security and Dependability, Karlsruhe Institute of Technology (KIT).}
 \email{lgschmid@kth.se}
 \orcid{0000-0002-3600-6899}
 \affiliation{%
   \institution{KTH Royal Institute of Technology}
   \city{Stockholm}
   \country{Sweden}
}

\author{Timur Sa\u{g}lam}
 \email{timur.saglam@kit.edu}
 \orcid{0000-0001-5983-4032}
 \affiliation{%
   \institution{KASTEL -- Institute of Information Security and Dependability, Karlsruhe Institute of Technology (KIT)}
   \city{Karlsruhe}
   \country{Germany}
}

\author{Nils Niehues}
 \email{nils.niehues@kit.edu}
 \orcid{0009-0006-4295-7232}
 \affiliation{%
   \institution{KASTEL -- Institute of Information Security and Dependability, Karlsruhe Institute of Technology (KIT)}
   \city{Karlsruhe}
   \country{Germany}
}

\renewcommand{\shortauthors}{Maisch et al.}

\begin{abstract}

Plagiarism detection in programming education faces growing challenges due to increasingly sophisticated obfuscation techniques, particularly automated refactoring-based attacks. While code plagiarism detection systems used in education practice are resilient against basic obfuscation, they struggle against structural modifications that preserve program behavior, especially caused by refactoring-based obfuscation. This paper presents a novel and extensible framework that enhances state-of-the-art detectors by leveraging code property graphs and graph transformations to counteract refactoring-based obfuscation. Our comprehensive evaluation of real-world student submissions, obfuscated using both algorithmic and AI-based obfuscation attacks, demonstrates a significant improvement in detecting plagiarized code.

\end{abstract}



\maketitle

\renewcommand{\subsectionautorefname}{Section}
\renewcommand{\subsubsectionautorefname}{Section}
\newcommand{\myname}{\ac{Nocte}} 

\section{Introduction}
Plagiarism is a prevalent challenge in computer science education, facilitated by the ease of duplicating and modifying digital assignments~\cite{Cosma2008, Murray2010, Le2013}. Addressing this challenge is crucial to maintain academic integrity.
Due to the high number of students in computer science courses, manual inspection is impractical~\cite{Camp2017, Kustanto2009}.
Moreover, students are creative in \textit{obfuscating} their plagiarism to hide the relation to the original source~\cite{Pawelczak2018}. In the case of programming assignments, students commonly utilize techniques such as renaming, reordering, or restructuring~\cite{Novak2019, novak2020, Karnalim2016}.

In light of these issues, educators commonly use software plagiarism detection systems~\cite{DevoreMcDonald2020}, which automate parts of the detection process and enable plagiarism detection at scale.
These detectors analyze sets of programs to identify pairs with a suspiciously high degree of similarity~\cite{prechelt2002}.
Most approaches compare the code structure~\cite{Nichols2019, Novak2019}, with token-based approaches like MOSS~\cite{MOSS} and JPlag~\cite{ prechelt2002} most widely used in practice~\cite{Aniceto2021, Novak2019, Lancaster2004}. 

Token-based approaches parse and linearize programs, capturing their structure in an internal representation.
On pairs of these linearized representations, matching code fragments of significant length are identified. The match coverage is then used to compute a similarity score and to derive suspicious candidates~\cite{SaglamDiss}. Nevertheless, assessing which candidates qualify as plagiarism is ultimately a human decision, given the underlying ethical considerations~\cite{Culwin2001, Weber2019}.
Considering only essential program elements, these approaches intentionally abstract from details such as names, types, and literals~\cite{prechelt2002}.
This makes these detectors inherently resilient against simple obfuscation attempts based on renaming, retyping, and other lexical changes~\cite{Joy1999}. 
However, for more advanced obfuscation attacks, their resilience remains insufficient~\cite{DevoreMcDonald2020, Luo2017, SaglamDiss}.

Despite these shortcomings, for novice programmers, evading detection is not feasible, as manually obfuscating a program is tedious and requires a profound understanding of programming languages ~\cite{Joy1999}. As long as defeating software plagiarism detectors takes more effort than completing the actual assignment, these detectors remain adequate~\cite{DevoreMcDonald2020} and essential in guiding educators' inspections of suspicious candidates~\cite{BottoTobar2022}.
However, this assumption has been broken with the recent rise of automated \textit{obfuscation attacks}~\cite{DevoreMcDonald2020, Foltynek2020, Biderman2022, Pawelczak2018}.
These \textit{obfuscation attacks} aim to avoid detection by automatically altering the structural properties of a program without changing its underlying behavior~\cite{Saglam2024b}.

Several recent works in the field of software plagiarism detection aim to design countermeasures against specific types of automated obfuscation attacks, such as the reordering of independent statements or the insertion of dead statements~\cite{Saglam2024b, Saglam2024a, Saglam2024c}.
However, existing approaches are not resilient against obfuscation attacks based on refactoring operations~\cite{eduard2025}.
\textit{Refactoring-based obfuscation attacks} modify programs via refactoring operations to change their structure while preserving their behavior to ensure plagiarism instances qualify as valid solutions.
In detail, they modify the program structure by applying refactoring operations at various permissible locations. These attacks include refactoring operations such as function extraction or inlining, introducing unnecessary control flow constructs, and restructuring loops and conditionals. Consequently, these attacks significantly alter the syntactic representation of programs, making it difficult to identify plagiarized code.
While early automated attacks relied solely on algorithmic approaches~\cite{DevoreMcDonald2020}, generative artificial intelligence can also be used to automate refactoring attacks~\cite{ChatGPTGuide, Daun2023}, making the obfuscation more accessible than ever before~\cite{Khalil_Er_2023, Saglam2024a}. Thus, both algorithmic and AI-based obfuscation attacks pose a viable threat to the detectors used in education practice.
Even attack-independent approaches~\cite{Saglam2025} do not provide sufficient resilience as they cannot systematically address complex program structure changes.

\subsubsection*{Approach}
This paper presents \acs{Nocte}, a framework to provide token-based plagiarism detection systems with resilience against refactoring-based obfuscation attacks.
\acs{Nocte} transforms programs into a normalized structure concerning semantically equivalent alternatives.
While token-based code plagiarism detectors usually tokenize the parsed program code directly, \acs{Nocte} transforms all input programs into \acp{CPG}~\cite{yamaguchi2014}. \acp{CPG}, typically used for software vulnerability analysis, combines a program's abstract syntax tree, evaluation order graph, and program dependence graph.
On these graphs, we apply a set of graph transformations specifically designed to counter obfuscation attacks. Next, we generate linear program representations from the normalized \acp{CPG} and pass them to the plagiarism detector, allowing it to proceed as usual.
To ensure a deterministic order, we employ topological sorting~\cite{kahn1962} during tokenization.
\acs{Nocte} is modular and extensible, as we separate the \ac{CPG}-based framework from the set of deobfuscation transformations. Thus, additional graph transformations can be integrated seamlessly to address emerging threats.

\subsubsection*{Evaluation}
%
We evaluate \acs{Nocte} based on the code plagiarism detector JPlag~\cite{prechelt2002, JPlag_GitHub}.
To that end, we employ four real-world datasets of student submissions and both algorithmic and AI-based obfuscation attacks.
We evaluate 1037 original and 540 obfuscated programs, encompassing more than 3.5 million pairwise comparisons.
%
The results show that \acs{Nocte} significantly outperforms the state-of-the-art for insertion-based and refactoring-based obfuscation attacks. While AI-based obfuscation is less reliable than algorithmic attacks, we show that it remains an insufficiently addressed challenge for \textit{all} current approaches.
Moreover, the results show that \acs{Nocte} enhances the detection of AI-generated programs despite not being designed for this task.
We provide our code and the evaluation data via the supplementary material~\cite{maisch2026nocte_supp}.

\subsubsection*{Contributions}
In this paper, we present three contributions:
\begin{enumerate}[label=\textbf{C\arabic*}]

    \item\label{c1} \acs{Nocte}, a \textit{\ac{CPG} transformation framework}, including a graph linearization and tokenization component, which allows to counter refactoring-based obfuscation attacks.
    \item\label{c2} A \textit{set of fourteen refactoring transformations} covering a wide variety of refactoring-based obfuscation attacks, thus providing resilience against such attacks.
    \item\label{c3} A comprehensive evaluation using real-world data sets with automatically obfuscated plagiarism instances based on both algorithmic and AI-based obfuscation.
\end{enumerate}
\section{Refactoring-based Obfuscation Attacks}\label{sec:example}

State-of-the-art detection approaches compare program structures by identifying similarities between code fragments~\cite{Nichols2019}.
Thus, obfuscation attacks aim to prevent the detector from matching related fragments by changing the program structure beyond trivial and lexical changes, such as renaming program elements.

\lstset{keepspaces=true}
\begin{table}
	\centering
	\small
	\setlength\tabcolsep{2pt}
    \caption{Original code and modified variant after inserting two statements (+), removing one ($-$), and altering two ($\sim$).}
    \hugTableCaption
	\label{tab:running-example}
	\begin{tabular}{clcl}
		\hline
		\#&\footnotesize\textbf{Original}                         & {$\to$}           & \footnotesize\textbf{Variant}                \\
		\hline
		1&\lstinline|printRoots(int n) {|                        &                   & \lstinline|printRoots(int n) {|              \\
		2&\lstinline|  |                                         & \cellcolor{add}\textbf{(+)}      & \cellcolor{add}\lstinline|  int i = 0;|      \\
		3&\cellcolor{mod}\lstinline|  for (int i=0; i<n; i++) {| & \cellcolor{mod}\textbf{($\sim$)} & \cellcolor{mod}\lstinline|  while (i < n) {| \\
		4&\lstinline|    double d = sqrt(i);|                    &                   & \lstinline|    double d = sqrt(i);|          \\
		5&\cellcolor{del}\lstinline|    d++;|                    & \cellcolor{del}\textbf{($-$)}    & \lstinline|    |                             \\
		6&\cellcolor{mod}\lstinline|    println(d);|             & \cellcolor{mod}\textbf{($\sim$)} & \cellcolor{mod}\lstinline|    println(++d);| \\
		7&\lstinline|    |                                       & \cellcolor{add}\textbf{(+)}      & \cellcolor{add}\lstinline|    i++;|          \\
		8&\lstinline|  }|                                        &                   & \lstinline|  }|                              \\
		9&\lstinline|}|                                          &                   & \lstinline|}|                                \\
		\hline
	\end{tabular}
    \figuresqueeze
\end{table}
\begin{table}
	\centering
	\small
    \caption{The two token sequences corresponding to the programs in \autoref{tab:running-example} with matching subsequences highlighted.}
    \hugTableCaption
	\label{tab:running-example-tokens}
	\begin{tabular}{cc c cc}
		\hline
		id               & \textbf{Original Tokens}         & $\to$             & id                 & \textbf{Variant Tokens}          \\
		\hline 
		\match
		1                & \texttt{method start}            &                   & 1                  & \texttt{method start}            \\
		\match
		2                & \texttt{variable}                &                   & 2                  & \texttt{variable}                \\

		                   &                                 & \textbf{(+)}      & \cellcolor{add}2   & \cellcolor{add}\texttt{variable} \\
		\match
		3                & \texttt{loop start}              &                   & 3                  & \texttt{loop start}              \\
		\match
		2                & \texttt{variable}                &                   & 2                  & \texttt{variable}                \\
		\cellcolor{del}4 & \cellcolor{del}  \texttt{assign} & \textbf{($\sim$)}      & \cellcolor{add}5   & \cellcolor{add}\texttt{apply}    \\
		\cellcolor{del}2 & \cellcolor{del}\texttt{variable} & \textbf{($-$)}      &                    &                                  \\
		\match 
		5                & \texttt{apply}                   &                   & 5                  & \texttt{apply}                   \\
		\match 
		4                & \texttt{assign}                  &                   & 4                  & \texttt{assign}                  \\
										         
		\cellcolor{del}5 & \cellcolor{del}\texttt{apply}    & \textbf{($\sim$)} & \cellcolor{add}4   & \cellcolor{add}\texttt{assign}   \\
		\match
		6                & \texttt{loop end}                &                   & 6                  & \texttt{loop end}                \\
		\match 
		7                & \texttt{method end}              &                   & 7                  & \texttt{method end}              \\
		\hline
	\end{tabular}
    \figuresqueeze
\end{table}

\autoref{tab:running-example} shows the effects of two specific refactoring operations on a 
program: First, the index-based \texttt{for} loop on line 3 is replaced with a \texttt{while} loop, requiring an increment statement on line 7. Second, the increment operation in line 5 is inlined into the print statement, changing it from a post-increment to a pre-increment operator.
Both resulting programs--the original and the variant--still print the square roots of a continuous sequence of integers. Yet, the two refactoring steps alter the program structure enough to evade detection through common software plagiarism detection systems.

Before we illustrate the effect of structural changes on the detection process, we look closer at the internal program representation used by such detection systems.
Plagiarism detectors used in education practice linearize programs by parsing them, traversing the parse tree, and capturing the type of relevant program elements as tokens~\cite{Saglam2024b}. The resulting token sequence is an abstract representation of the program structure, discarding details such as names, data types, and values. In essence, they can be regarded as finite sequences of natural numbers~\cite{SaglamDiss}.
\autoref{tab:running-example-tokens} depicts the token sequences of the two programs in \autoref{tab:running-example}. Note that the \textit{id} shows the numerical representation.
Plagiarism detectors compute the similarity between program pairs by identifying matching subsequences within their corresponding token sequences.
To ensure that matches are meaningful, these detectors define a minimal match length below which subsequences are ignored. While this is crucial to avoid false positives, this threshold also introduces a vulnerability for obfuscation:
By altering the original program code, obfuscation attacks attempt to fragment matching token subsequences into smaller parts until they fall below the set matching threshold and are thus discarded.
In fact, for \textit{any} obfuscation attack to be effective, it needs to affect the token sequence. 

Returning to the example, the token sequences of both programs are different due to the modifications that created the variant from the original program. Let us assume the minimal match length threshold is 3.
We can identify four matching subsequences, highlighted in gray, where each neighboring pair of matches is interrupted by unmatched tokens.
Due to these unmatched tokens, all matches fall below the minimum match length threshold. Thus, the plagiarism detector would discard these matches, resulting in a similarity of \qty{0}{\percent} for these two programs.

The concepts illustrated based on the example shown in \autoref{tab:running-example}, even though demonstrated at a rather small scale, also apply to larger programs.
Moreover, the refactoring operations used here are two typical examples among many possible refactoring operations that can be effectively exploited as obfuscation attacks.
Other possible operations include the extraction and inlining of methods, replacing \texttt{if} cascades with \texttt{switch} statements, and creating dummy variables, methods, or classes~\cite{Karnalim2016}.

\begin{figure}
    \centering
    \includegraphics[width=\linewidth]{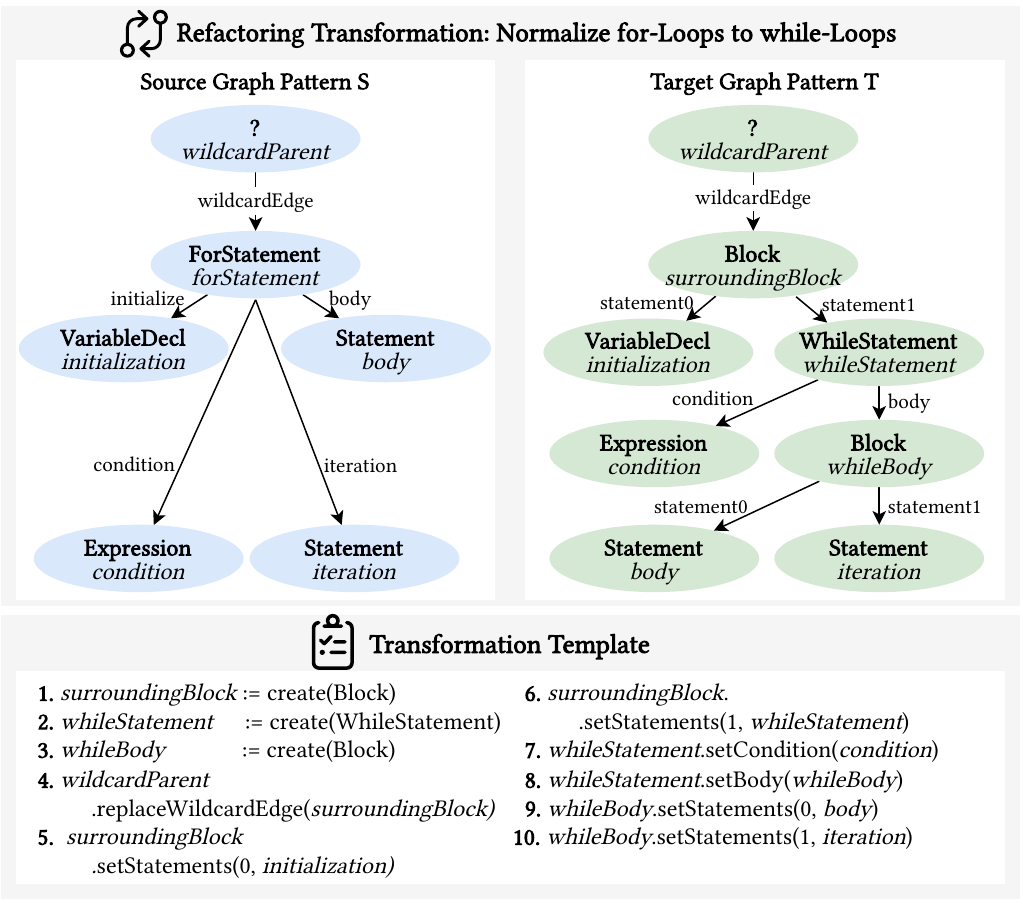}
    \hugPlotCaption
    \caption{Example of a refactoring transformation with a corresponding transformation template containing operations on how to refactor the source graph pattern $S$ into the target graph pattern $T$. Each node pattern shows its type (bold) and its role (italicized).}
    \label{fig:transformation-preparation}
    \figuresqueeze
\end{figure}
\section{Approach}\label{sec:approach}

As established in the previous section (cf. Section~\ref{sec:example}), refactorings are an effective method to obfuscate plagiarism: They alter the code structure, thereby breaking up matches and reducing the similarity of the resulting token sequence when compared to the original. 
To mitigate the effects of refactoring-based obfuscation attacks on token-based plagiarism detection, we introduce the \acf{Nocte}, a code transformation framework that provides structural normalization for code submissions as a preprocessing step for token-based plagiarism detection (\contribution{1}). 
The transformations that \ac{Nocte} employs for the normalization are modular and easily adaptable. We provide an initial set of transformations (\contribution{2}) which address refactorings frequently referenced in the literature~\cite{Karnalim2016, novak2016}.

\begin{table}[t]
	\centering
	\small
    \caption{Normalized version of both the original and variant code of \autoref{tab:running-example}, illustrated here as code instead of a CPG. 
    }
    \hugTableCaption
	\label{tab:running-example-normalized}
	\small
    \begin{tabular}{cl}
		\hline
		\#&\footnotesize\textbf{Normalized Version} \\
		\hline
		1&\lstinline|printRoots(int n) {    | \\
		2&\lstinline|  int i = 0;           | \\
		3&\lstinline|  while (i < n) {      | \\
		4&\lstinline|    println(sqrt(i++) + 1); | \\
		5&\lstinline|  }                    | \\
		6&\lstinline|}                      | \\
		\hline
	\end{tabular}
    \tablesqueeze
\end{table}

The core idea of our approach is as follows: Instead of directly linearizing the program into a token sequence, as is common for token-based plagiarism detectors, we first construct a graph-based representation for each program.
On these graphs, we can apply graph transformations to normalize their structures and thus reverse the effects of obfuscation attacks. For this purpose, we employ \acfp{CPG} due to their expressiveness and detailed program representation.
After executing the graph transformations, we linearize the graph representations and pass them to the plagiarism detector, allowing the detection process to proceed as usual.

Returning to the example submissions illustrated in \autoref{tab:running-example}, for example, we could realign both versions by replacing \textit{all} \texttt{for} loops with \texttt{while} loops and inlining single-use assignments. As the resulting \textit{normalized} code structures of both versions (\autoref{tab:running-example-normalized}) are identical, their corresponding token sequences are identical as well. Thus, the plagiarism detectors will easily identify this pair as highly similar, raising the attention of instructors, and successfully overcoming the initial obfuscation attempt.

\begin{figure}
    \centering
    \includegraphics[width=\linewidth]{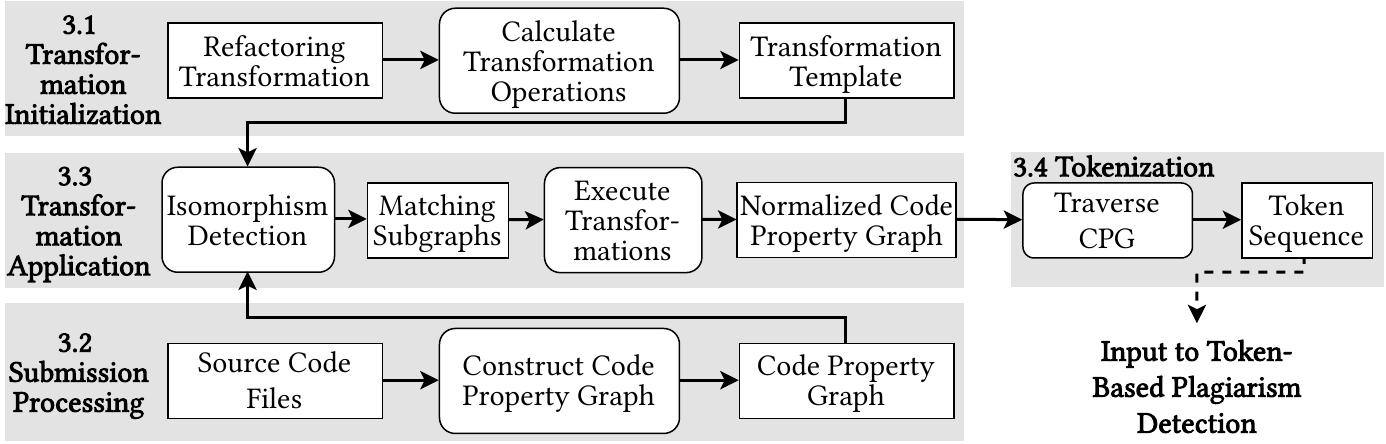}
    \hugCaption
    \caption{Pipeline of our Normalization Procedure.}
    \label{fig:pipeline}
    \figuresqueeze
\end{figure}

In detail, \acs{Nocte} incorporates four steps, as illustrated in \autoref{fig:pipeline}:
As a one-time initialization, each transformation is analyzed to derive its individual transformation operations from its textual representation (\autoref{ssec:transformationInitialization}).
For each submission, a \ac{CPG} is constructed first, capturing detailed syntactic and semantic information about the code structure (\autoref{ssec:submissionProcessing}).
Then, each transformation is applied to the CPG until it reaches a fixed, normalized state (\autoref{ssec:transformationApplication}).   
Finally, the normalized CPG is traversed to extract a token sequence for the submission (\autoref{ssec:tokenization}). This token sequence is the output of the transformation engine and the basis for pairwise comparison of submissions by the plagiarism detection tool.

Once integrated into the user's token-based plagiarism detection system of choice, the entire normalization process is a fully automated step in the existing pipeline, requiring no further user interaction.

The following sections describe each step of the pipeline in detail.

\subsection{Transformation Initialization}
\label{ssec:transformationInitialization}
We represent graph transformations as pairs $(S, T)$ of a source and target graph pattern, adapted from~\cite{mens2005}, shown in \autoref{fig:transformation-preparation} (top). First, we determine the individual \textit{transformation operations} needed to perform each transformation, i.e., to adapt the structure of $S$ to $T$. To this end, we compare $S$ and $T$ using in-parallel depth-first search, selecting the appropriate operation to address each difference between the two patterns. The resulting sequence of operations $ops$, shown in \autoref{fig:transformation-preparation} (bottom), and the source graph $S$ constitute the \textit{transformation template} for a transformation. 

\subsection{Submission Processing}
\label{ssec:submissionProcessing}
For each code submission in the corpus, a \ac{CPG} is constructed, the intermediate representation on which we will perform the normalization step. This \ac{CPG} generation entails parsing the submission into an \ac{AST} and then iteratively extending the tree with additional edges and node attributes, capturing in-depth structural and semantic connections across the graph. The finished \ac{CPG} combines the \ac{AST} with an \ac{EOG}, a \ac{DDG}, and a \ac{CDG}. We base our \ac{CPG} construction pipeline on the \ac{CPG} framework presented by Fraunhofer AISEC~\cite{weiss2022}.

\subsection{Transformation Application}
\label{ssec:transformationApplication}
The \ac{CPG} $\mathcal{G}$ is normalized using the given set of transformations. To this end, each transformation $\mathcal{T}$ is iteratively applied at every substructure of $\mathcal{G}$ possible. After that, the process continues with the subsequent transformation. The individual steps of the transformation application are described in more detail in the following paragraphs.

\subsubsection{\textbf{Isomorphism detection.}}
For each transformation template $(S,ops)$, a graph isomorphism detection algorithm iteratively identifies subgraphs of the submission \ac{CPG} $\mathcal{G}=(N,V)$ on which the transformation can be applied, i.e., subgraphs which match the structure of the source graph pattern $S$. Many transformations specify additional conditions that a matching subgraph must satisfy to ensure the transformation will not change the program behavior. 
Each \textit{match} of $S$ in $\mathcal{G}$ stores the mapping of concrete \ac{CPG} nodes of the matching subgraph to the \textit{role} that they assume in the course of the transformation.

\subsubsection{\textbf{Transformation instantiation and execution.}}
\label{ssec:transformationExecution}
For each match of $\mathcal{T}$, the transformation operations $ops$ are carried out on the concrete \ac{CPG} nodes which correspond to the different roles of the transformation. 
After all matches are transformed, \ac{Nocte} searches for matches with the same transformation template again, as new matching subgraphs may have emerged as a result of the previous transformations, where the transformation can be applied. 

\subsubsection{\textbf{Normalization of statement order}}
State-of-the-art token-based plagiarism detection is already resilient against reordering of large code elements\repl{, e.g.}{ such as} methods; however, at the statement level, obfuscations via permutation and insertion remain effective. To defend against these obfuscation attacks, we use the control flow information of the \ac{EOG} subgraph to determine data dependencies between method statements. Then, we deterministically sort the statements, taking into account their mutual dependencies. Also, statements with no transitive data dependencies to the program state (\textit{dead code}) are removed from the \ac{CPG}, including not only the declaration of unused variables, but also unused assignments.

This normalization step cannot be expressed as a set of \ac{CPG} transformations, as it does not target a specific structure in the code. Rather, it is applied as an intermediate step between two phases of transformation.

\subsection{Tokenization}
\label{ssec:tokenization}
\subsubsection{Linearization} To derive a linear token sequence out of an inherently non-linear code graph, we have to specify the order in which we want to traverse various elements on various levels, e.g., the members of a class, the subexpressions of a statement. While the state-of-the-art approach uses preorder depth-first search to replicate the \ac{AST} structure, we opt for postorder depth-first search, which resembles the \ac{EOG} structure---the order in which subexpressions may be evaluated at runtime. In particular cases, this eliminates the opportunity to use simple syntactical variants for plagiarism obfuscation, e.g., defining a \texttt{for} loop variable in the \texttt{for} statement vs. before the \texttt{for} statement.

\subsubsection{Token Selection} \label{sssec:tokenSelection} 
To allow for seamless integration of \ac{Nocte} into an existing token-based plagiarism detection system, we may directly reuse the existing tokenization rules, thereby replicating the abstraction level and the specific language feature selection that will be represented in the token sequence.

\subsection{Limitations}\label{sec:limits}

This section discusses the limitations and potential areas for extension of the presented normalization.

\renewcommand*{\descriptionlabel}[1]{\hspace\labelsep{\normalfont\sffamily\itshape#1:}}
\begin{description}[style=unboxed,leftmargin=0cm]
\item[Language dependence] While our \ac{CPG} approach supports multiple programming languages, the refactoring transformations are designed for Java code. Some of them are applicable in related languages like C\#, Python, or C++; however, testing for wide language support was outside of the scope of our evaluation. We expect that language-specific refactoring transformations must be tailored to each programming language to ensure effectiveness. 
\item[Selection of transformations] \acs{Nocte} is designed to counter precisely the set of refactoring-based obfuscation attacks covered by the supplied refactoring transformations, providing tailored, modular defense against specific attacks. 
This means that attacks not part of the refactoring transformations may still be effective. 
We propose refactoring transformations that cover common refactoring attacks (cf. Section~\ref{sec:definition-transformations}) and provide easy ways to extend the list of refactoring transformations.

\item[Conservative semantic-preserving approach] To avoid an increased false-positive rate, we ensure that each step of the normalization pipeline does not alter the semantics of the code. Adversaries, on the other hand, may employ semantic-changing obfuscation attacks, which our approach does not address.
However, these types of attacks are uncommon~\cite{SaglamDiss}, as semantically deviant programs typically do not pass as correct solutions.
Even if semantic-changing obfuscation is employed, we expect that \acs{Nocte} will still help distinguish plagiarism from unrelated programs.
\end{description}

\section{Definition of Refactoring Transformations} \label{sec:definition-transformations}
In this section, we present our second main contribution (\contribution{2}), the selection
of transformations: We selected 13 refactoring strategies from the literature~\cite{Karnalim2016, novak2020}
and designed 14 corresponding transformations to counter them. 
Our goal was a diverse set of attacks with a wide range of complexity and scope, broadly applicable to programs and thus easy to automate. 
Large-scale refactoring operations, such as the introduction of design patterns, are less broadly applicable as they must be tailored to the program at hand and require expertise to be applied correctly; thus, we consider large-scale refactoring out of scope.
Instead, we focus on broadly applicable operations, ranging in granularity from statement-level to class-level transformations.
We group the attack schemes in four categories: \begin{enumerate*}[label=(\roman*)]
    \item\textit{Inserting Elements},
    \item\textit{Moving Elements},
    \item\textit{Extracting Elements}, and
    \item\textit{Semantically Equivalent Replacement}.
\end{enumerate*} 

In the next sections, each category of refactoring attack and the corresponding normalizing transformations will be described in more detail.

\subsection{Inserting Elements} 

The insertion of elements may split matching token subsequences, or dilute the overall similarity by adding large amounts of code that do not contribute to the program behavior. According to \citet{novak2020}, many of these elements are considered \textit{common code} and may be removed in the context of plagiarism detection. Also, empty classes may result from other transformations.
We support the removal of these inserted elements with the following transformations:

{
    \renewcommand*{\descriptionlabel}[1]{\hspace\labelsep{\normalfont\sffamily\itshape#1:}}
    \begin{description}[style=unboxed,leftmargin=0pt]
        \item[(1) Remove Empty Methods] void-type methods with an empty block are removed.
        \item[(2) Remove Empty Constructors] constructors with an empty block are removed.
        \item[(3) Remove Empty Classes] classes with no methods, fields, or inner classes are removed.
        \item[(4) Getter Methods] non-void methods that directly return either a field of the surrounding class or a constant are removed.
        \item[(5) Unsupported Methods] methods that immediately throw an unconditional exception are removed.
        \item[(6) Unsupported Constructors] constructors that immediately throw an unconditional exception are removed.
    \end{description}
}
Since unsupported methods and constructors are not part of the intended program behavior, we treat them as inserted elements and remove them. 

\subsection{Moving Elements} 

Another attack scheme is to move code fragments away from their immediate context, according to a usage analysis. A strategy we have often observed in real-life submissions is the creation of dedicated utility classes, where class constants and auxiliary methods are moved to obfuscate plagiarism.
We support the analysis and context restoration for the following elements:
{
    \renewcommand*{\descriptionlabel}[1]{\hspace\labelsep{\normalfont\sffamily\itshape#1:}}
   \begin{description}[style=unboxed,leftmargin=0pt]
        \item[(7) Move Constants To Only Using Class] Constants which are used only in one particular class, different from the current declaring class of the constant, are moved into that class.
    \end{description}
}
\subsection{Extracting elements} 

Refactoring-based obfuscation attacks in this category may create additional declarations and/or function calls by extracting or wrapping code elements.
We support inlining the following elements:
{
    \renewcommand*{\descriptionlabel}[1]{\hspace\labelsep{\normalfont\sffamily\itshape#1:}}
    \begin{description}[style=unboxed,leftmargin=0pt]
        \item[(8) Inline Single-use Variables] Local variables are inlined if (i) they are referenced exactly once after their declaration, (ii) that reference is a read access, and (iii) the expression assigned to it in its declaration does not change value until that reference. 
        \item[(9) Inline Single-use Constants] Class constants are inlined if they are referenced exactly once. In contrast to single-use variables, we expect constants to be initialized with a constant value.
        \item[(10/11) Inline Optional Values] Values wrapped in an Optional object which handles the case that the value is \texttt{null}. This comprises a transformation that replaces the creation of the wrapped value with the value itself, and another that replaces the unwrapping method call with the proper value.  
    \end{description}
}
The transformation \textit{Inline Single-use Variable} preserves the semantics of the code only if the value of the assigned expression remains unchanged up to the potential inlining position. To verify this precondition, we use the \ac{CPG}'s data flow information.

\subsection{Semantically Equivalent Replacement} 

By replacing specific structures by a semantically equivalent other structure in the source code, the token sequence of a plagiarism instance can be changed to obfuscate its relation to the original. We consider the following replacements: 

{
    \renewcommand*{\descriptionlabel}[1]{\hspace\labelsep{\normalfont\sffamily\itshape#1:}}
    \begin{description}[style=unboxed,leftmargin=0pt]
        \item[(12) Revert Negated If-Else] If the condition expression of an \texttt{if-else} statement is a negation expression, the \texttt{then} and \texttt{else} blocks are swapped, and the negated inner expression replaces the condition.
        \item[(13) Revert If-Unequal-Else] If the condition expression of an \texttt{if-else} statement is an inequality expression, the \texttt{then} and \texttt{else} blocks are swapped, and an equality expression replaces the condition.
        \item[(14) For Loop To While Loop] All \texttt{for} loops are replaced by the equivalent \texttt{while} statement, moving the declaration of the loop variable before the \texttt{while} statement, and the iteration statement to the end of the \texttt{while} block.
    \end{description}
}
To ensure the preservation of the semantics through the transformation \textit{For Loop To While Loop}, the scope of the newly created loop variable must end after the \texttt{while} statement. To that end, the \texttt{while} statement and the definition of its loop variable must be moved into an inner block.
There are countless variants of possible refactorings involving if-else blocks, e.g., the inclusion of the optional \texttt{else} block if the \texttt{if} block always ends in a \texttt{return statement}, and also notably, the usage of a condition which is constant at runtime, but complex enough that conservative static analysis is unable to determine that it is constant.

\section{Evaluation}\label{sec:eval}
In this section, we present the evaluation of \ac{Nocte} (\contribution{3}).

We employ four real-world datasets and generate plagiarism instances with four types of real-world automated obfuscation attacks. 
In total, we used 1037 original and 540 obfuscated programs, resulting in over 3.5 million pairwise comparisons.
The results demonstrate that \ac{Nocte} not only matches the resilience of other approaches but also significantly outperforms them for refactoring-based and insertion-based obfuscation attacks.
We provide our code and the evaluation data via the supplementary material~\cite{maisch2026nocte_supp}.

\subsection{Methodology}\label{ssec:methodology}

With our evaluation, we set out to answer the following questions:
\begin{description}[style=unboxed,leftmargin=5pt]\label{eval-q}
    \small
    \item[Q1] Does \ac{Nocte} provide resilience to insertion-based attacks?
    \item[Q2] Does \ac{Nocte} provide resilience to \add{mixed} refactoring-based attacks?
    \item[Q3] Does \ac{Nocte} provide resilience to \add{mixed} AI-based attacks?
    \item[Q4] Does \ac{Nocte} improve the detection of AI-generated programs?
\end{description}
We choose the plagiarism detector JPlag~\cite{prechelt2002, JPlag_GitHub} as a baseline as it is widely used, frequently referenced in literature, and open-source~\cite{Aniceto2021, Novak2019}. We also compare \ac{Nocte} with two related approaches from our earlier work: \begin{itemize}
    \item \ac{TSN}~\cite{Saglam2024b}, a pre-comparison optimization approach on the token sequences; and 
    \item \ac{SMM}~\cite{Saglam2025}, a post-comparison optimization approach on the matches.
\end{itemize}
To our knowledge, these are the only state-of-the-art, tool-independent approaches that target \textit{automated} obfuscation in the context of plagiarism detection. Note that while clone detection approaches are related, they do not consider attacker-defender scenarios and are thus tampering-prone (see \autoref{sec:rw}).
\ac{TSN} enables normalization based on program dependence graphs, thus providing resilience against insertion-based and reordering-based obfuscation. \ac{SMM} is an algorithm that heuristically merges interrupted code fragments for plagiarism detectors as post-processing. Thus, it provides resilience against a variety of obfuscation attacks.
\ac{Nocte}, \ac{TSN}, and \ac{SMM} are all implemented based on JPlag and can thus be directly compared to the baseline.
\todoRobin{For CR: These are *our* old approaches.}

For all evaluation questions, we employ the following metrics:
\begin{description}[style=unboxed,leftmargin=5pt]\label{eval-m}
    \small
    \item[M1] Separation between values plagiarism instances and\\ unrelated programs (median similarity difference).
    \item[M2] Statistical significance of the results compared to the baseline\\ (based on one-sided Wilcoxon signed-rank tests).
    \item[M3] Practical significance of the results compared to the baseline\\ (based on Cliff's delta~\cite{Cliff1993} and its interpretation~\cite{Romano2006}).
\end{description}
As educators use plagiarism detectors to identify suspicious outliers, \mref{1} measures how well the plagiarism instances are separated from unrelated programs. To test if the approaches improve the baseline JPlag, \mref{2} and \mref{3} evaluate for statistical and practical significance via statistical tests.

We now discuss the obfuscation attacks and datasets we used.

\subsubsection{Obfuscation Attacks}\label{par:evalObfuscationAttacks}
We generate plagiarism instances from existing programs by applying three different automated obfuscation attacks: dead code insertion, refactoring, and AI-based code rewriting. We omit statement-level reordering-based obfuscation, as it is not an effective obfuscation attack~\cite{Saglam2024b}.
Moreover, we evaluate the similarity of programs fully AI-generated based on the assignment description, which does not constitute obfuscation in a classical sense. 
In the following, we will only \textit{briefly} discuss these obfuscation attacks, as we do not want to encourage employing these attacks.
 
For insertion-based obfuscation, we utilize the attack outlined by \citet{DevoreMcDonald2020} and implemented for Java code by \citet{Saglam2024b}. This attack inserts dead statements from the original program and a pool of pre-defined statements. After each insertion, the resulting program is compiled to ensure syntactical correctness and preservation of program behavior.

For refactoring-based obfuscation, we leverage \textit{Spoon}~\cite{Pawlak2006} and automatically apply multiple behavior-preserving refactoring operations at random positions at the \ac{AST} level to obfuscate a given program, simulating an automated combined obfuscation attack pattern.
While this only includes refactoring operations that \ac{Nocte} can address, we do not evaluate the effect of individual refactoring types in isolation; instead, we assess the impact of repeatedly applying all refactoring types in combination, ensuring that our evaluation focuses on the overall obfuscation effect.
In detail, the refactoring operations include optional wrapping, extracting expressions as new variables, introducing constant container classes and extracting constants, swapping if-else statements and inverting the corresponding conditions, inserting methods and constructors, and introducing access methods for existing fields.

To complement the two algorithmic obfuscation attacks, we employ an AI-based attack using OpenAI's GPT-4 as a third strategy. We use 16 different zero-shot prompts to obfuscate pre-existing programs, mimicking how students might use generative AI to obfuscate their plagiarism.
This selection of prompts results from a preliminary study in which we evaluated nearly 50 prompts with varying degrees of specificity. Some prompts specified obfuscation techniques (e.g., \textit{insert dead code}), while others were formulated in a deliberately general style (\textit{modify the code to look different but behave as before}). The study revealed that more specific prompts are less effective in modifying structural properties of the programs, thus resulting in high similarities between the original and obfuscated program; consequently, we restrict our evaluation to rather unrestrictive, more effective prompts---listed in~\cite{maisch2026nocte_supp}---which range from requesting minor structural changes to requesting a refactored version of the original program. Note that there is no guarantee that the program behavior is preserved by GPT-4.

For the final evaluation stage, we generate complete programs from the textual assignment description.

\subsubsection{Datasets}
For our evaluation, we use two tasks from the publicly available PROGpedia dataset~\cite{paiva2023} and the datasets \textit{TicTacToe} and \textit{BoardGame} from our prior work~\cite{Saglam2024b}.
In total, we use four real-world datasets of programs created by university students, originating from different courses and assignment types (see \autoref{tab:prog-datasets}).
We filter out manual, human-made plagiarism across all datasets to evaluate automated obfuscation.
\autoref{tab:plagiate} shows the number of plagiarism instances generated from these datasets.

We use two tasks from the \textit{PROGpedia} dataset~\cite{paiva2023}.
Task 19 involves the design of a graph data structure and a depth-first search to analyze a social network, while
Task 56 concerns minimum spanning trees using Prim's algorithm. Both datasets contain small to medium-sized Java programs, and incorrect solutions were omitted.
Additionally, we use two datasets from mandatory assignments of an introductory programming course, \textit{TicTacToe} and \textit{BoardGame}. These datasets contain command-line-based Java implementations of the paper-and-pencil game Tic-Tac-Toe and a comprehensive board game, respectively. \textit{TicTacToe} includes 626 medium-sized programs, while \textit{BoardGame} contains 434 very large programs.

\begin{table}
    \centering
    \caption[Evaluation Datasets]{Evaluation datasets, their number of programs, mean lines of code (LOC), programming language, and source.}
    \label{tab:prog-datasets}
    \hugTableCaption
    \small
    \begin{tabular}{lrrcc}
        \toprule
        {Dataset Name} & {Programs} & {Mean LOC} & Language & Source \\
        \midrule
        PROGpedia19 & 27 & 131 & Java & \cite{paiva2023}\\
        PROGpedia56 & 28 & 85 & Java & \cite{paiva2023}\\
        TicTacToe & 626 & 236 & Java & \cite{Saglam2024b}\\
        BoardGame & 434 & 1529 & Java & \cite{Saglam2024b}\\
        \bottomrule
    \end{tabular}
    \figuresqueeze
\end{table}

\begin{table}
	\centering
	\small
    \caption[Obfuscation Attacks]{Overview of the number of plagiarized programs per dataset and obfuscation attack type (540 programs in total).}
    \hugTableCaption
	\label{tab:plagiate}
	\begin{tabular}{lcccc}
        \hline
		Attack Type             & Pp19 & Pp56 & TTT & BG        \\
        \hline
		Insertion-based Obf.    & 27        & 28     & 50  & 20        \\
		Refactoring-based Obf.  & 27        & 28     & 50  & 20        \\
		AI-based Obf.           & 80        & 80     & 80  & -         \\
		AI-based Generation     & -         & -      & 50  & -         \\
        \hline
	\end{tabular}
    \figuresqueeze
\end{table}

\subsection{Results}
In this section, we present the evaluation results. We discuss each obfuscation attack separately.
For each approach, we compare the pairwise similarity values of plagiarism pairs (ideally high) with those of unrelated program pairs (ideally low). The greater their separation, the easier it is for educators to spot plagiarism cases (cf. \mref{1}). To show statistical and practical significance (cf. \mref{2} and \mref{3}), we conduct statistical tests, as illustrated in \autoref{tab:plags} and \autoref{tab:originals}.

\subsubsection{Insertion-based Obfuscation}
\begin{figure*}[ht]
    \centering
    \centering
    \includegraphics[width=\textwidth,trim={0 10pt 0 0}]{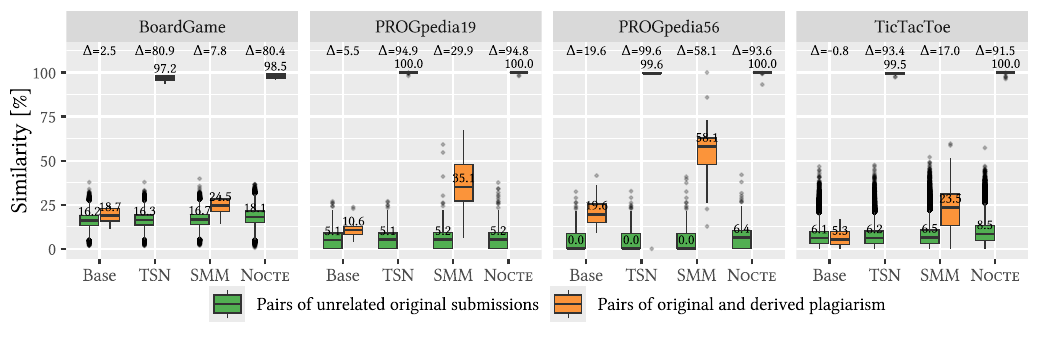}
    \groupHugPlotCaption
    \caption{Similarities for unrelated human programs and plagiarism instances based on insertion-based obfuscation.}
    \label{fig:evalInsertion}
    \plotsqueeze
\end{figure*}

First, we examine the resilience against obfuscation via dead statement insertion~(\qref{1}).
\autoref{fig:evalInsertion} shows the results for each dataset and each approach, as well as the unmodified version of JPlag as the baseline. The plots show all median similarity values and the difference between them.
For insertion-based obfuscation, \ac{Nocte} and \ac{TSN} provide the strongest separation between plagiarism pairs and unrelated programs, with median differences of 80--\qty{99}{percentage~points~(pp)}. Both approaches achieve near-immunity against insertion-based obfuscation. While \ac{SMM} also increases separation compared to the baseline, the effect is less pronounced, with median differences of 7.8--\qty{58.1}{pp}. Thus, some overlap between plagiarism pairs and unrelated programs remains for \ac{SMM}.
The statistical tests (\autoref{tab:plags}) show that all three approaches are statistically and practically significant improvements over the baseline. Notably, both \ac{Nocte} and \ac{TSN} exhibit a \textit{very large} effect size for all datasets.
\summaryBox{1}{\ac{Nocte} provides strong resilience against in\-ser\-tion-based obfuscation, outperforming \ac{SMM}, and matching \ac{TSN}.}

\subsubsection{Refactoring-based Obfuscation} \label{sec:eval:refactoring}
\begin{figure*}[ht]
    \centering
    \includegraphics[width=\textwidth,trim={0 10pt 0 0},clip]{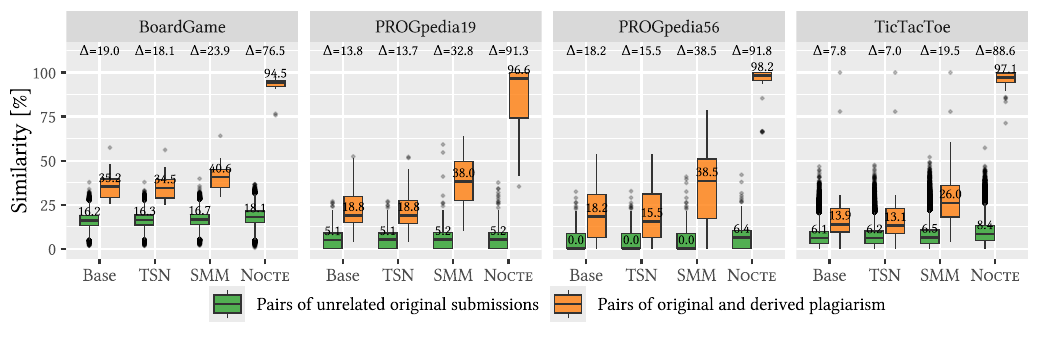}
    \groupHugPlotCaption
    \caption{Similarities for unrelated human programs and plagiarism instances based on refactoring-based obfuscation.}
    \label{fig:evalRefactoring}
    \plotsqueeze
\end{figure*}

Then, we examine resilience against refactoring-based obfuscation (\qref{2}).
\autoref{fig:evalRefactoring} shows the corresponding results.
For refactoring-based obfuscation, \ac{Nocte} outperforms the other approaches and is the only one to achieve strong separation between plagiarism pairs and unrelated programs with median differences of 76--\qty{92}{percentage~points~(pp)}, thus achieving near-immunity.
\ac{TSN} offers no benefit over the baseline, as it is designed only to provide resilience against insertion- and reordering-based obfuscation.
While \ac{SMM} provides improved separation over the baseline, the effect is less pronounced, with median differences of 23--\qty{39}{pp}. Again, some overlap remains for \ac{SMM}.
The statistical tests (\autoref{tab:plags}) show that the improvements of \ac{Nocte} and \ac{SMM} are both statistically and practically significant. For \ac{Nocte}, the effect size is \textit{very large}, while for SMM, it is only \textit{medium} to \textit{large}.

\summaryBox{2}{\ac{Nocte} provides strong resilience against re\-fac\-tor\-ing-based obfuscation, significantly outperforming both \ac{SMM} and \ac{TSN}.}

\begin{figure*}[ht]
    \centering
    \includegraphics[width=\textwidth,trim={0 10pt 0 0}]{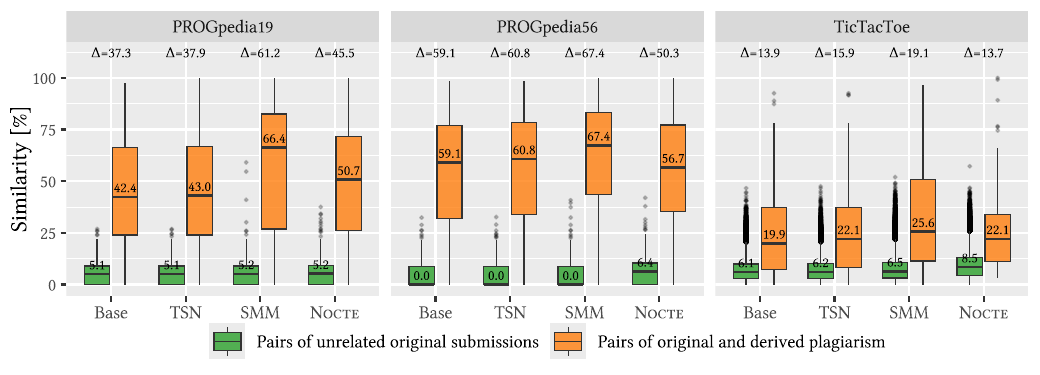}
    \groupHugPlotCaption
    \caption{Similarities for unrelated human programs and plagiarism instances based on AI-based obfuscation.}
    \label{fig:evalGptObfuscation}
    \plotsqueeze
\end{figure*}

\subsubsection{AI-based Obfuscation}
Subsequently, we examine resilience against AI-based obfuscation (\qref{3}). Here, we omit the \textit{BoardGame} dataset because the programs exceed GPT-4's token limit. Note that GPT-4 offers no guarantee of preserving the original program behavior, even when instructed to do so in the prompt. This is unlike algorithmic obfuscation, which ensures semantic preservation by design.
\autoref{fig:evalGptObfuscation} shows the corresponding results.
Compared to algorithmic obfuscation, the variance of the similarity values is high. Furthermore, AI-based obfuscation is not particularly strong, as the overlap between plagiarism and original programs is limited.
However, all approaches show only limited improvements.
Again, \ac{TSN} shows no improvement over the baseline. \ac{SMM} and \ac{Nocte} show limited improvement, with \ac{SMM} showing the best results. However, no approach achieves clear separation.
The statistical tests (\autoref{tab:plags}) show statistical significance for \ac{Nocte} and \ac{SMM} (except for PROGpedia56 for \ac{Nocte}). However, practical significance is \textit{negligible} to small for both.
\summaryBox{3}{Although AI-based obfuscation is less reliable than other attacks, \ac{Nocte} currently has no significant impact on resilience against it.}

\subsubsection{AI-generated Programs}\label{sec:eval:aiGeneration}
\begin{figure}[ht]
    \centering
    \includegraphics[width=\linewidth,trim={0 10pt 0 0}]{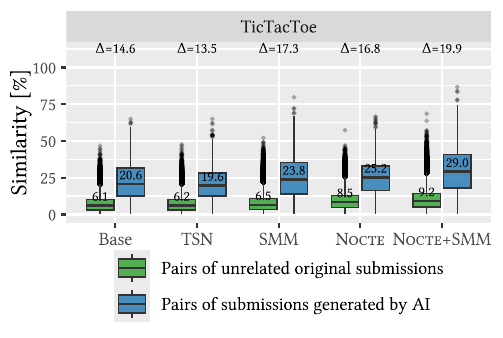}
    \hugPlotCaption
    \caption{Similarities for unrelated human programs (green) and AI-generated programs (orange).}
    \label{fig:evalGptGeneration}
    \plotsqueeze
\end{figure}

Finally, we examine the similarity of AI-generated programs (\qref{4}). We only use one dataset here, as the full assignment description is required. Note that AI-based code generation does not constitute plagiarism in a classical sense~\cite{Novak2019, Saglam2024a}. Thus, not all approaches are designed for their detection.
\autoref{fig:evalGptGeneration} shows the results comparing the similarity of unrelated human programs with the similarity among AI-generated programs. Note that even for the baseline, AI-generated programs exhibit a higher similarity than human-made ones.
Among the approaches, \ac{Nocte} and \ac{SMM} provide the largest improvement with a separation of around \qty{17}{pp}. \ac{Nocte} shows the highest median similarity for a single approach of around 25 percent.
Notably, combining \ac{Nocte} with \ac{SMM} achieves even better results. Here, the separation reaches around \qty{20}{pp}, and the generated programs exhibit a median similarity of 29 percent. 
The statistical tests (\autoref{tab:plags}) show that the improvement compared to the baseline is statistically significant for \ac{Nocte}, \ac{SMM}, and the combination of both. The practical significance is limited for either approach, but their combination reaches a measurable, albeit \textit{low}, effect size.

\summaryBox{4}{\ac{Nocte} improves the detection of AI-generated programs, especially when combined with \ac{SMM}.}

\subsection{Discussion}\label{ssec:discussion}
In this section, we discuss our findings and their implications.

\paragraph{Algorithmic Obfuscation.}
The evaluation results for the algorithmic obfuscation attacks---insertion-based and refactoring-based---show that \ac{Nocte} outperforms the state of the art. The normalization provided by \ac{Nocte} achieves near-optimal separation between plagiarism instances and unrelated original programs, reverting virtually all effects of the obfuscation. The results show that existing approaches lack resilience to refactoring-based obfuscation; \ac{Nocte} overcomes this limitation.
Since the defense mechanisms for insertion-based and refactoring-based obfuscation are separate steps in the pipeline of \ac{Nocte}, a combined obfuscation attack using both strategies is no more effective than either attack alone.

\begin{table}
	\centering
	\footnotesize
	\caption{One-sided Wilcoxon signed-rank test comparing improvements over the baseline ($\alpha=0.01$, $H1=greater$), with effect size (Cliff's $\delta$), interpretation ($\delta Int.$), and \qty{95}{\percent} confidence interval ($CI$). Low $p$ and high $\delta$ are ideal.}
    \hugTableCaption
	\label{tab:plags}
	\begin{tabular}{crrrrrrr}
        \hline
        Obf.                                                          & ds                    & Approach  & $p$     & $\delta$ & $\delta\,$Int. & $\delta$ \qty{95}{\percent} CI     \\ 
		\hline
		\multirow{12}{*}{\rotatebox[origin=c]{90}{Insertion-based}}   & \multirow{3}{*}{BG} & TSN       & 4.8e-05 & 1        & Very Large     & [ 1.00,  1.00]       \\ 
		                                                              &                       & SMM       & 4.8e-05 & 0.56     & Large          & [ 0.22,  0.78]       \\ 
                                                                    &                       & \myname{}     & 1.1e-04 & 1        & Very Large     & [ 0.99,  1.00]       \\                        
		\cline{2-7}
		                                                              & \multirow{3}{*}{Pp19} & TSN       & 3e-06   & 1        & Very Large     & [ 1.00,  1.00]       \\ 
		                                                              &                       & SMM       & 2.2e-05 & 0.82     & Very Large     & [ 0.55,  0.93]       \\ 
                                                                    &                       & \myname{}     & 3e-06   & 1        & Very Large     & [ 1.00,  1.00]       \\ 
		\cline{2-7}
		                                                              & \multirow{3}{*}{Pp56} & TSN       & 2.2e-06 & 0.93     & Very Large     & [ 0.63,  0.99]       \\ 
		                                                              &                       & SMM       & 4.4e-06 & 0.88     & Very Large     & [ 0.66,  0.96]       \\ 
                                                                    &                       & \myname{}     & 2e-06   & 1        & Very Large     & [ 1.00,  1.00]       \\ 
		\cline{2-7}
		                                                              & \multirow{3}{*}{TTT}  & TSN       & 3.9e-10 & 1        & Very Large     & [ 1.00,  1.00]       \\ 
		                                                              &                       & SMM       & 2.7e-09 & 0.79     & Very Large     & [ 0.63,  0.89]       \\ 
                                                                    &                       & \myname{}     & 3.8e-10 & 1        & Very Large     & [ 1.00,  1.00]       \\ 
		\hline
		\multirow{12}{*}{\rotatebox[origin=c]{90}{Refactoring-based}} & \multirow{3}{*}{BG} & TSN       & 1       & -0.06    & Negligible     & [-0.39,  0.29]      \\ 
		                                                              &                       & SMM       & 4.8e-05 & 0.42     & Medium         & [ 0.05,  0.69]       \\ 
                                                                    &                       & \myname{}     & 1.6e-04 & 1        & Very Large     & [ 0.99,  1.00]       \\ 
		\cline{2-7}
		                                                              & \multirow{3}{*}{Pp19} & TSN       & 1       & -0.04    & Negligible     & [-0.33,  0.26]      \\ 
		                                                              &                       & SMM       & 1.4e-05 & 0.53     & Large          & [ 0.23,  0.74]       \\ 
                                                                    &                       & \myname{}     & 3e-06   & 0.98     & Very Large     & [ 0.91,  0.99]       \\ 
		\cline{2-7}
		                                                              & \multirow{3}{*}{Pp56} & TSN       & 0.31    & -0.01    & Negligible     & [-0.31,  0.28]      \\
		                                                              &                       & SMM       & 1.6e-04 & 0.35     & Medium         & [ 0.04,  0.59]       \\ 
                                                                    &                       & \myname{}     & 2e-06   & 1        & Very Large     & [ 1.00,  1.00]       \\ 
		\cline{2-7}
		                                                              & \multirow{3}{*}{TTT}  & TSN       & 0.98    & -0.02    & Negligible     & [-0.24,  0.21]      \\ 
		                                                              &                       & SMM       & 2.7e-09 & 0.51     & Large          & [ 0.29,  0.68]       \\ 
                                                                    &                       & \myname{}     & 5.7e-10 & 0.96     & Very Large     & [ 0.81,  0.99]       \\ 
		\hline
		\multirow{9}{*}{\rotatebox[origin=c]{90}{AI-based}}         & \multirow{3}{*}{Pp19} & TSN       & 0.26    & 0.01     & Negligible     & [-0.14,  0.16]      \\ 
		                                                              &                       & SMM       & < 1e-10 & 0.19     & Small          & [ 0.03,  0.34]       \\ 
                                                                    &                       & \myname{}     & 1.8e-10 & 0.09     & Negligible     & [-0.06,  0.24]      \\ 
		\cline{2-7}
		                                                              & \multirow{3}{*}{Pp56} & TSN       & 0.016   & 0.05     & Negligible     & [-0.10,  0.20]      \\ 
		                                                              &                       & SMM       & < 1e-10 & 0.2      & Small          & [ 0.05,  0.35]       \\ 
                                                                    &                     & \myname{}     & 0.067   & 0.04     & Negligible     & [-0.12,  0.19]      \\ 
		\cline{2-7}
		                                                              & \multirow{3}{*}{TTT}  & TSN       & 0.012   & 0.04     & Negligible     & [-0.12,  0.19]      \\ 
		                                                              &                       & SMM       & < 1e-10 & 0.11     & Negligible     & [-0.04,  0.26]      \\ 
                                                                    &                     & \myname{}     & 3.1e-05 & 0.06     & Negligible     & [-0.09,  0.21]      \\ 
		\hline
		\multirow{4}{*}{\rotatebox[origin=c]{90}{AI Gen.}}           & \multirow{4}{*}{TTT}  & TSN       & 1       & -0.07    & Negligible     & [-0.12, -0.03]    \\ 
		                                                              &                       & SMM       & < 1e-10 & 0.1      & Negligible     & [ 0.05,  0.15]      \\ 
                                                                    &                       & \myname{}   & < 1e-10 & 0.13     & Negligible     & [ 0.09,  0.18]      \\ 
		                                                              &                       & \myname{}+SMM & < 1e-10 & 0.29     & Small          & [ 0.25,  0.33]      \\ 
		\hline
	\end{tabular}
    \figuresqueeze
\end{table}


\begin{table}
	\centering
	\footnotesize
    \caption{One-sided Wilcoxon signed-rank test comparing potential adverse effects on false-positives over the baseline ($\alpha=0.01$, $H1=greater$), with effect size (Cliff's $\delta$), interpretation ($\delta\,$Int.), and \qty{95}{\percent} confidence interval ($CI$). High $p$ and low $\delta$ are ideal.}
    \hugTableCaption
	\label{tab:originals}
	\begin{tabular}{crrrrrr}
		\hline
		Dataset                                              & Approach  & $p$     & $\delta$ & $\delta\,$Int. & $\delta$ \qty{95}{\percent} CI        \\ 
		\hline
		\multirow{3}{*}{\rotatebox[origin=c]{0}{BoardGame}}   & TSN       & < 1e-10 & 0.02     & Negligible     & [ 0.01, 0.03]      \\ 
                                                              & SMM       & < 1e-10 & 0.07     & Negligible     & [ 0.07, 0.08]       \\ 
                                                              & \myname{}     & < 1e-10 & 0.24     & Small          & [ 0.24, 0.25]       \\ 
		\hline
		\multirow{3}{*}{\rotatebox[origin=c]{0}{PROGpedia19}} & TSN       & 8.3e-05 & 0        & Negligible     & [-0.08, 0.09]        \\ 
                                                              & SMM       & 0.011   & 0.02     & Negligible     & [-0.07, 0.10]        \\ 
                                                              & \myname{}     & 0.0098  & 0.03     & Negligible     & [-0.05, 0.12]        \\ 
		\hline
		\multirow{3}{*}{\rotatebox[origin=c]{0}{PROGpedia56}} & TSN       & 0.23    & -0.03    & Negligible     & [-0.11, 0.04]       \\ 
                                                              & SMM       & 0.018   & 0.02     & Negligible     & [-0.06, 0.09]        \\ 
                                                              & \myname{}     & 3.4e-08 & 0.14     & Negligible     & [ 0.06, 0.21]         \\ 
		\hline
		\multirow{4}{*}{\rotatebox[origin=c]{0}{TicTacToe}}  & TSN       & < 1e-10 & 0.01     & Negligible     & [ 0.01, 0.01]     \\ 
                                                             & SMM       & < 1e-10 & 0.05     & Negligible     & [ 0.04, 0.05]     \\ 
                                                             & \myname{}     & < 1e-10 & 0.23     & Small          & [ 0.22, 0.23]     \\ 
                                                             & \myname{}+SMM & < 1e-10 & 0.28     & Small          & [ 0.28, 0.29]     \\ 
		\hline
	\end{tabular}
    \figuresqueeze
\end{table}


The mechanisms used to automatically obfuscate the submissions bear no direct relation to the defense mechanisms of \ac{Nocte}: Insertion-based obfuscation is a real-world attack~\cite{DevoreMcDonald2020} which is straightforward to implement but requires data flow analysis to reverse. For the refactoring-based obfuscation, the refactorings employed conceptually map to the counter-transformations, but are implemented independently: The obfuscation is based on the code-level \textit{Spoon} meta-programming library for code transformation, while \ac{Nocte} operates on the graph level via the CPG.

\paragraph{AI-based Obfuscation.}\label{par:discussionAIObfuscation}
We observe that AI-based obfuscation is less effective than algorithmic obfuscation. Even for the baseline, there is a limited overlap between plagiarism and originals, especially for the PROGpedia datasets (median similarity for PROGpedia56: Insertion: \qty{19.6}{\percent}, Refactoring: \qty{18.2}{\percent}, AI-based plagiarism: \qty{60.8}{\percent}, see Figs. \ref{fig:evalInsertion}, \ref{fig:evalRefactoring}, \ref{fig:evalGptObfuscation}). This high baseline similarity of AI-obfuscated programs limits improvement across all approaches. 

The high variance in similarity suggests that AI-based obfuscation is inconsistently efficient and thus less reliable. Students do not know the overall similarity distribution, so they do not know what obfuscation level is sufficient to evade detection.
Moreover, AI-based obfuscation does not guarantee the preservation of the original program behavior.

Despite these limitations of AI-based obfuscation, our results indicate that AI-based obfuscation attacks remain an insufficiently addressed challenge in plagiarism detection.
No approach we evaluate provides a notable improvement over the baseline.
Evidently, further research is required to adapt these approaches for AI-based obfuscation. 
\ac{TSN} only addresses insertion-based and reordering-based obfuscation and is thus conceptually limited.
While \ac{SMM} operates heuristically, providing limited resilience for this obfuscation attack, it offers no inherent room for further improvement.
For \ac{Nocte}, the current resilience against AI-based obfuscation is also limited. 
The results indicate that some of the modifications by GPT-4 are currently not covered by our set of transformations. \repl{GPT-4 applies a wide variety of modifications to obfuscate programs.}{
In an analysis of the generated programs, we identified some of these additional obfuscation strategies, such as (a) loop changes, (b) algorithm changes, and (c) usage of API function calls. 
We can easily extend \ac{Nocte} for additional refactorings, e.g., (a), whereas reimplementation techniques and semantic clones, such as (b) and (c), are out of scope for \ac{Nocte}.
Nonetheless, in the light of the variety of modifications which GPT\nobreakdash-4 used to obfuscate programs,} further analysis of these plagiarism instances is required to extend the transformation set and thereby strengthen the resilience of \ac{Nocte} against AI-based obfuscation.

\paragraph{AI-generated Programs.}\label{par:discussionAIGeneration}
Although \ac{Nocte} was not designed to detect AI-generated programs, it notably improves the similarity of generated programs and thus increases the chance of their detection. For better results, we recommend using it in combination with SMM.
Interestingly, programs generated with GPT-4 exhibit higher similarity among each other than human ones.
This may be due to the inherent determinism of large language models~\cite{SaglamDiss}.
This is the case across all analyzed approaches.
However, it is crucial to note that software plagiarism detectors are not the ideal solution for detecting AI-generated code. Current research points towards signature- or watermark-based methods, for example, based on recognizing specific patterns or characteristics inherent to AI-generated content~\cite{zhao2024provable, Jiang2023}.
It should also be noted that cheating via AI-generated programs is only viable if the generated programs are valid. Currently, we observe that AI-generated programs work reliably only for small-scale tasks; however, they exhibit a higher structural similarity than programs created by humans (see~\autoref{sec:eval:aiGeneration}). To exploit this, instructors can add AI-generated programs to the submission set for the plagiarism check~\cite{pang2024}. To address the rapid developments of generative AI, future research is necessary.

\paragraph{Emerging Threats}
To address emerging threats, additional defense mechanisms are needed for both AI-based obfuscation and AI-generated code.
As discussed, signature- or watermark-based methods remain promising options.
Due to recent advances, another direction is to incorporate AI itself into detection techniques.
However, several challenges remain:
1) limited explainability, which is crucial for ethical transparency; 2) the need for traceability beyond similarity scores, such as fragment matching and visualization, which vector embeddings, for example, lack; 3) scarce real-world educational data, increasing the risk of overfitting; and 4) ethical concerns, as ML-based methods may produce false positives; LLMs, for example, are known to be prone to hallucination.

Despite these challenges, exploring the incorporation of AI remains a promising endeavor.

\paragraph{Impact on False Positives.}
Plagiarism detectors must achieve a low false-positive rate to be ethical and viable solutions for educators. Thus, we also analyzed the impact of all approaches on the unrelated, original programs.
\autoref{tab:originals} shows the result of the statistical tests regarding a potential adverse impact on these programs compared to the baseline. Note that we are testing for an increase in similarity relative to the original value; Thus, a high p-value and low effect size are desirable.
While all approaches show statistically significant effects on unrelated programs in some datasets, the effect sizes show that the effects are practically insignificant:
\ac{SMM} and \ac{Nocte} have small to \textit{negligible} effect sizes;
\ac{TSN} has \textit{negligible} effect sizes.
Any normalization technique is inherently bound to have \textit{some} effect on all programs; nevertheless, the median similarity differences also show that the similarity increases for plagiarism pairs outweigh those for unrelated pairs.

\paragraph{Impact on Performance.}
On a consumer notebook (Ryzen 7 Pro, 64GB RAM), we measured the runtimes for all datasets (10 runs each). \autoref{tab:performance} shows the results. 
As plagiarism detection occurs only a few times per semester, these low runtimes are negligible.
\begin{table}
    \centering
    \small
    \ifthenelse{\boolean{showdiff}}{\color{blue}}{}
    \caption{Runtime of 10 runs of \ac{Nocte} for each dataset.}
    \hugTableCaption
    \begin{tabular}{lccccc}
        \toprule
        \textbf{Task} & \textbf{Mean time} & \textbf{Std.dev.} & \textit{\textbf{n}} & \textbf{Total LOC}\\\midrule
         BoardGame & 09:20 min & 00:06 min & 407 & 660 KLOC	\\
         PROGpedia19 & 00:33 min & 00:02 min & 161 & 25 KLOC \\
         PROGpedia56 & 00:24 min & 00:01 min & 164 & 17 KLOC \\
         TicTacToe & 03:44 min & 00:08 min & 845 & 200 KLOC \\\bottomrule
    \end{tabular}
    \label{tab:performance}
    \figuresqueeze
\end{table}

\subsection{Threats to Validity}\label{ssec:ttv}
In the following section, we discuss how we address threats to validity as outlined by \citet{Wohlin2012}, \citet{runeson2008}.
\paragraph{Internal Validity.} 
To ensure internal validity, we use JPlag as a foundation for all approaches and keep all other factors unchanged. The original programs on which the automatic obfuscation attacks are applied to create plagiarized instances are chosen randomly from each dataset to avoid cherry-picking. For AI-based obfuscation, we use 16 prompts derived through systematic \textit{prompt engineering} to achieve consistently reproducible results. While the refactoring-based obfuscation mechanism and the normalization engine cover conceptually equivalent refactorings, both were designed independently and using different technologies to ensure meaningful results.
\paragraph{External Validity.} \label{par:extValidity}
To ensure external validity, we base our evaluation datasets on real-world student submissions from multiple sources, each with varying corpus sizes, submission lengths, code complexity, and assignment characteristics. While \ac{Nocte} was integrated into JPlag for our evaluation, \add{its conceptual design is independent from JPlag; thus, }it can be incorporated into any other token-based plagiarism detection system \add{such as \textsc{Moss} and Dolos}. Furthermore, it is possible to combine \ac{Nocte} with other optimization approaches, as demonstrated with \ac{SMM}.

While the selected refactoring transformations (\contribution{2}) are listed as common in reference literature, they do not cover all strategies students may use to obfuscate plagiarized programs. A thorough study of real student plagiarism cases may address this gap.

\paragraph{Construct Validity.} 
To ensure construct validity, we employ the methodological standards of prior software plagiarism detection research. This is reflected in the fact that we carefully prepared the labeled datasets, used a GQM plan~\cite{Basili1984, Basili1992}, selected the evaluation metrics, and established an approach-independent ground truth.  
\paragraph{Reliability.} For reliability, we provide the code for our approach and the evaluation data as supplementary material~\cite{maisch2026nocte_supp}.

\section{Related Work}\label{sec:rw}
In this section, we discuss various related works.

\paragraph{Plagiarism Detection and Obfuscation Resilience.}
\citet{Saglam2024b} use PDGs as an intermediate representation to detect dead code and counter the reordering of independent statements. While \myname{} also addresses these obfuscation attacks, it also counters refactoring-based attacks. 
\citet{Karnalim2016} uses tokens to represent Java Bytecode instructions instead of structural code elements. This approach is immune to textual changes like renaming, but also against semantic-preserving instruction replacement. However, it is conceptually limited to JVM-based languages. 
\citet{Saglam2025} counter various obfuscation attacks by heuristically combining adjacent token matches that are separated only by a few tokens.
While this provides broad resilience, it is less suited against complex, refactoring-based obfuscation that \myname{} targets (Section~\ref{sec:eval:refactoring}). 
\citet{maisch2025tolerant} introduce tolerant token matching, an alternative approach to dealing with syntactic variety by accepting different, but conceptually similar token types as matches. While this approach is effective, it faces the problem of interdependent syntactic structures; e.g., \texttt{for} loops usually contain a loop variable declaration, whereas \texttt{while} loop variables must be declared before the loop. \myname{}'s structural normalization solves this problem. 
\citet{novak2016} proposes preprocessing code by, e.g., removing comments and common code to emphasize more original fragments of code.
We build on these ideas by extracting 14 concrete refactorings (Section~\ref{sec:definition-transformations}) and integrating them into a flexible framework.
Moreover, our approach does not modify the code itself, as it operates on CPGs and normalizes the internal code representation of plagiarism detectors.

In addition to token-based approaches, there are some graph-based approaches to plagiarism detection~\cite{Novak2019, ferrante1987, liu2006, cheers2021}. While graph-based approaches may be resilient against some obfuscation attacks like dead code insertion, they are not feasible in practice~\cite{liu2006} due to the $\mathcal{NP}$ nature of subgraph isomorphism~\cite{Shang2008, McCreesh2020, Lubiw1981}.
Our approach leverages the potential of graph-based techniques while preserving the scalability of token-based techniques.

\paragraph{Automated Refactorings.}
There is extensive research on automated refactoring of source code~\cite{baqais2020automatic}. 
\citet{liu2024-3erefactor} perform refactorings on code to achieve architectural consistency. 
\citet{lin2016-refactorings} apply design patterns to code via automated refactorings. 
\citet{alizadeh2020} propose an interactive tool to take developers' feedback on proposed refactorings into account. 
These approaches refactor software to make it more understandable and support the development process, and thus have to respect the needs of developers~\cite{szoke2016-designing}, such as providing understandable transformations. 
In contrast, \myname{} applies refactorings to normalize code to the same representations to enable assessment of code pairs. 

\paragraph{Code Property Graphs.}
We also relate to research on CPGs beyond plagiarism detection.
\citet{yamaguchi2014} introduce CPGs for software vulnerability analysis, modeling common vulnerabilities as templates that can be detected by traversing a program's \ac{CPG}. 
Based on this, neural networks have been used to learn these vulnerable patterns in CPGs~\cite{zhou2019neurips, Chakraborty2022}. Other approaches aim to detect vulnerabilities using large language models~\cite{lu2024}. 
These approaches use CPGs for security analysis, in contrast to our use case of plagiarism detection. 

\paragraph{Software Clone Detection.}
Software systems often contain similar fragments called code clones~\cite{Roy2009}, which impede modern software development~\cite{juergens2009}. Clone detection has been extensively researched~\cite{Shobha2021review, ain2019, rattan2013}.
While it is similar to plagiarism detection~\cite{Novak2019}, they ultimately differ~\cite{mariani2012}.
Clone detection seeks similarities in a single program, while plagiarism detection compares sets of programs.
Moreover, clones are created accidentally~\cite{juergens2009}, while plagiarism is a deliberate act.
Most importantly, clone detectors are insufficient for plagiarism detection because they do not account for adversarial scenarios~\cite{SaglamDiss}. Thus, they are vulnerable to obfuscation attacks~\cite{Saglam2024a}.

\add{However, some works in code clone detection specifically create a normalization of code before performing the similarity detection, thus relating to our work.
\citet{davis2010} propose normalizing code through intermediate representations to abstract away non-significant syntactic differences. To that end, they analyze compiler-generated assembler code.
However, this approach relies on binaries and faces challenges due to platform-specific compilation artifacts.
\citet{kononenko2014} show in a study that clone detection results can differ substantially between source code and Java byte code, especially in large systems.
\citet{selim2010} improve the detection of Type 3 clones (\textit{near-miss clones})~\cite{rattan2013}  with a hybrid approach that analyzes both Java source code and compiler-generated intermediate representations. 
The approach benefits from the compiler normalization but is inherently Java-specific.
\citet{ragkhitwetsagul2017} propose decompiling Java bytecode back into source code to allow existing Java-based clone detectors to identify clones that may be obscured in the original source. While this pre-processing step could be applied for plagiarism detection, it would obscure idiosyncrasies used as evidence during misconduct investigations.
In contrast, our approach can be implemented for other languages and is platform-independent. Moreover, it only normalizes the program's internal representation, thus preserving traceability and explainability.
}

\section{Conclusion}\label{sec:conclusion}
This paper introduces \acs{Nocte}, a \ac{CPG}-based framework to enhance token-based plagiarism detectors against refactoring-based obfuscation attacks (\contribution{1}). Our approach significantly improves obfuscation resilience by applying targeted graph transformations (\contribution{2}) before tokenization while maintaining a minimal impact on unrelated programs.
Our evaluation (\contribution{3}) is based on real-world datasets. 
The results show that \acs{Nocte} significantly outperforms the state of the art for insertion-based and refactoring-based obfuscation attacks.
As code obfuscation techniques continue to evolve, our framework provides an extensible solution to strengthen academic integrity in programming education.
As future work, a controlled study where students use AI tools to plagiarize given programs could provide deeper insight into obfuscation strategies beyond our current selection.
\begin{acks}
    This work was supported by funding from the pilot program Core Informatics at KIT (KiKIT) of the Helmholtz Association (HGF) and the BMBF (German Federal Ministry of Education and Research) grant number 16KISA086 (ANYMOS).
\end{acks}

\bibliographystyle{ACM-Reference-Format}
\bibliography{bibliography}

\end{document}